\documentclass[letter]{ptptex}

\usepackage{graphicx}
\newcommand{\beq}{\begin{equation}}
\newcommand{\beqa}{\begin{eqnarray}}
\newcommand{\eeq}{\end{equation}}
\newcommand{\eeqa}{\end{eqnarray}}
\newcommand{\siml}{\lesssim}
\newcommand{\simg}{\gtrsim}
\notypesetlogo                       %comment in if to eliminate PTPTeX 
\title{%        %You can use \\ for explicit line-break.
Gamma-Ray Burst without Baryonic and Magnetic Load?
}

\author{%       %Use \scshape for the family name.
Kunihito \textsc{Ioka},
Yutaka \textsc{Ohira},
Norita \textsc{Kawanaka}
and Akira \textsc{Mizuta}
}

\inst{%     %Affiliation, neglected when [addenda] or [errata].
KEK Theory Center
and Graduate University for Advanced Studies (Sokendai), 
Tsukuba 305-0801, Japan
}

\abst{%         %This abstract is neglected when [addenda] or [errata].
We show that, contrary to common belief, 
internal shocks can arise in an accelerating radiation-dominated jet
if it is confined even weakly
to a converging opening angle
because the acceleration declines.
The radiation-dominated internal shock (RDIS)
enables a very efficient yet highly nonthermal emission
by Fermi-like photon acceleration,
keeping the electron-positron ($e^{\pm}$) pair
photosphere 
and inertia up to a high Lorentz factor $>1000$.
In gamma-ray bursts (GRBs), a weak confinement would
persist beyond the progenitor star 
or surrounding matter
because of the fast cocoon accompanying the breakout jet.
The simplest model predicts few high-energy cosmic rays and neutrinos, and
a correlation between the early afterglow and the GeV-TeV prompt emission.
The central engine allows a less fine-tuned baryon load
than previously thought,
even including
pure-leptonic unmagnetized outflows.
}

\begin{document}

\maketitle

\noindent {\it \S GRB and baryon load} --
Gamma-Ray Bursts (GRBs) are the most luminous objects in the universe.
It remains a big challenge to
reveal how most of the energy can be converted
into gamma rays with highly nonthermal spectra.
\cite{Meszaros:2006rc,Zhang:2007nka}.

The baryon load is a key parameter
for the fireball dynamics and emission.
The original fireball model \cite{p86,g86} was pure leptonic 
with the photon-to-baryon ratio $\eta\equiv L/{\dot M} c^2=\infty$.
As the fireball expands, the $e^{\pm}$ pairs annihilate
and almost all the energy is released to thermal radiation:
it is efficient but inconsistent with nonthermal observations.
This leads to the baryon-loaded model with $\eta \siml 10^3$,
in which the radiation energy is converted 
to the kinetic energy of matter, \cite{sp90}
and then back to the radiation by internal shocks
within variable outflows. 
\cite{Rees:1987,Rees:1994nw,Kobayashi:1997jk}
By its very nature, however, this standard picture
faces an efficiency problem. \cite{Zhang:2005fa,Ioka:2005zj}
The Lorentz factor dispersion required for
the efficient reconversion
is too large to reproduce the observed spectral correlations
$\nu_{\rm peak} \propto L^{1/2}$ 
\cite{Amati:2002ny,Yonetoku:2003gi,Ghirlanda:2009de}.
This brings forth the photosphere model
\cite{Thompson:1994,Rees:2004gt,Pe'er:2003ft,Ioka:2007qk,Ryde:2009wn,Ioka:2010xc}
that invokes a partial thermalization near the photosphere
to stabilize the peak energy $\nu_{\rm peak}$ at the thermal peak,
while retaining the nonthermal spectrum.
The observed Band spectrum could be reproduced
by Comptonization of the thermal photons
via dissipation such as 
shocks \cite{Beloborodov:2009be,Ioka:2010xc,Lazzati:2010af}
and magnetic reconnections. \cite{Thompson:1994,Giannios:2007yj}

\begin{figure}[t]
%\centerline{\includegraphics[scale=.35]{f1.eps}}
\centerline{\includegraphics[scale=0.3]{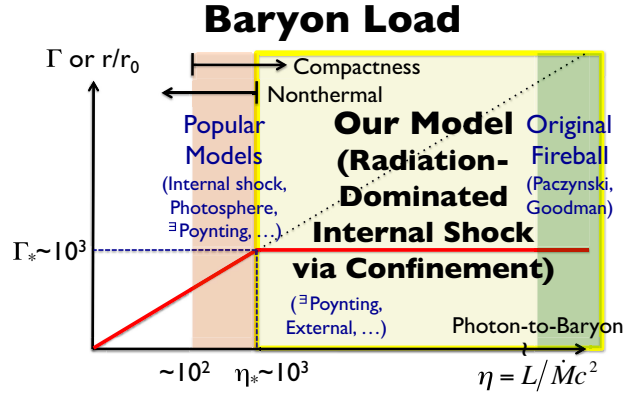}}
\caption{
We categorize GRB models on the basis of the baryon load.
Popular models usually assume
the photon-to-baryon ratio $\eta=L/{\dot M}c^2$
from $10^2 \siml \eta$ (to avoid the compactness problem)
to $\eta \siml 10^3$.
We consider the radiation-dominated jet with $\eta \simg 10^3$,
including the pure-leptonic original fireball model \cite{p86,g86}.
Such jets were considered to cause no internal shocks
before being transparent at the Lorentz factor $\Gamma \sim 10^3$
and radius $r/r_0 \sim 10^3$,
and hence, emit (dominant) thermal radiation
unlike in observations. \cite{Meszaros:1999gb}
%This picture is changed by the confinement of the jet.
This picture is changed by the jet confinement.
}
\label{fig:baryon}
\vspace{-.3cm}
\end{figure}

These matter-dominated models assume the baryon load in the range (see Fig.~\ref{fig:baryon})
\beqa
10^2 \siml \eta \left(\equiv \frac{L}{{\dot M} c^2}\right)
\siml 10^3,
\label{eq:eta}
\eeqa
to avoid the compactness problem \cite{Meszaros:2006rc}
(for $10^2 \siml \eta$)
and the predominant thermal emission \cite{Meszaros:1999gb}
(for $\eta \siml 10^3$),
where the baryon-poor fireball with $\eta \simg 10^3$ 
is considered to emit thermal photons
because no internal shocks occur
in the radiation-dominated fireball,
which accelerates as $\Gamma \propto r$ with
the preceding shell always faster than the succeeding one.
Therefore, the inferred baryon load is only limited 
within one order of magnitude,
in contrast to a huge variety of GRBs.
We may identify it as the ``{\it fine-tuning problem of the baryon load} ''.
Although the baryon-rich fireballs could be just unobservable
with selection bias,
there seems no compelling reason to believe $\eta \siml 10^3$.
Recent Fermi observations \cite{Abdo:2009a,Ackermann:2010,Zhang:2010ey}
have made the problem even worse \cite{Ioka:2010xc}
by detecting high-energy photons to
limit the Lorentz factor $\Gamma \simg 10^3$, i.e., 
$\eta \simg 10^3$ in some bursts
(despite the controversies on the numerical factor 
\cite{Zhao:2010kd,Zou:2010xg,Hascoet:2011vp}).

One possible solution to the fine-tuned baryon load problem
is the Poynting-dominated model. 
\cite{Thompson:1994,Meszaros:1996ww,Spruit:2000zm,Giannios:2007yj,Komissarov:2008ic,Tchekhovskoy:2009mf}
Even without (electrons associated with) baryon,
the energy is released only nonthermally via magnetic reconnections.
Many uncertainties still remain, especially in the dissipation site.
It may also be difficult to keep the jet cold in the hot photon pool
near the central engine.

The other possibility may be the externally induced dissipation 
by the oblique shocks \cite{Aloy:1999ai,Levinson:2006br,Bromberg:2007kw}
or the entrainment of baryon \cite{Ioka:2010xc}
during the acceleration.
However, the external dissipation only occurs
at the causally connected periphery of the jet,
and the strong confinement ($\theta < \Gamma^{-1}$)
is necessary for the full dissipation across the jet.

In this paper, we show that, contrary to popular belief, 
the radiation-dominated jet can cause internal shocks
if the geometry is confined even weakly
(even $\theta > \Gamma^{-1}$)
by some external pressure or internal magnetic field.
This could markedly expand the available parameter space to $10^3 \siml \eta$
as outlined below.

~

\begin{figure}[t]
%\centerline{\includegraphics[scale=1.3]{f1.eps}}
\centerline{\includegraphics[scale=1.3]{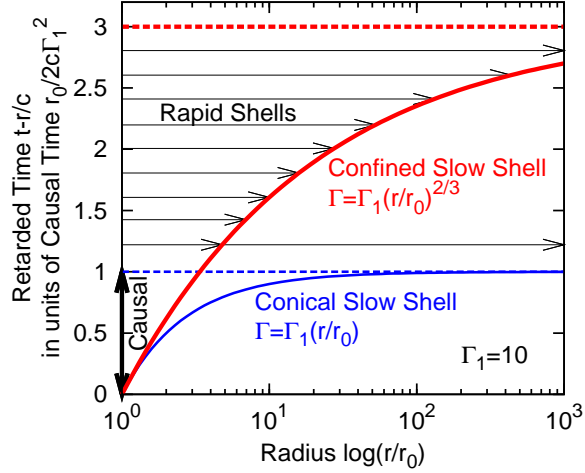}}
\caption{
A spacetime diagram for the retarded time $t-r/c$
in units of the causal time $r_0/2 c \Gamma_1^2$
as a function of the radius $\log (r/r_0)$,
using $\Gamma \gg 1$.
A photon travels horizontally in this diagram.
A conical radiation-dominated jet $\Gamma \propto r$ (solid line)
approaches a light path,
so that any rapid jet emitted after 
the causal time $\sim r_0/2c\Gamma_1^2$
cannot catch up with it.
Meanwhile, a confined jet accelerates more slowly
$\Gamma \propto r^{\lambda}$
(thick solid line) in
a converging opening angle $\theta \propto r^{\lambda-1}$
with $\lambda<1$,
enabling internal shocks with rapid jets (arrows)
over a wide radius range.
%Here we assume the relativistic case $\Gamma \gg 1$.
}
\label{fig:spacetime}
\end{figure}

\noindent {\it \S Two shell model} --
We envisage that the Lorentz factor $\Gamma$ of an outflow fluctuates
by a factor of $\Delta \Gamma\sim$ a few around its mean.
This may be modeled by two shells
ejected from a radius $r_0$ at times $t_1$ and $t_2=t_1+\Delta t$
(where $\Delta t>0$)
with different initial Lorentz factors $\Gamma_1$ and $\Gamma_2 (>\Gamma_1)$,
respectively
(allowing for the possibility of $\Gamma_i \gg 1$ as in the jets 
breaking out of the progenitor star).
The shell trajectory is given by
\beqa
\frac{dr}{dt}=c\beta=c\sqrt{1-\frac{1}{\Gamma^2}}.
\label{eq:drdt}
\eeqa
For the conical (or spherical) 
radiation-dominated jet $\Gamma \propto r$,
we can easily integrate Eq.~(\ref{eq:drdt}) as
$r=\{({r_0^2}/{\Gamma_i^2})+\left[c(t-t_i)+r_0 \beta_i\right]^2\}^{1/2}$,
$(i=1, 2)$.
Two shells collide at
\beqa
\frac{r}{r_0}=\frac{1}{{\cal R}}\left[
1
+\left(\frac{1}{2\Gamma_{1}^2}+\frac{1}{2\Gamma_2^2}\right) {\cal R}^2
+\left(\frac{1}{4\Gamma_1^2}-\frac{1}{4\Gamma_2^2}\right)^2 {\cal R}^4
\right]^{1/2},
\eeqa
if the following condition is satisfied,
\beqa
{\cal R}\equiv 
\frac{2}{\beta_2^2-\beta_1^2} 
\left(\beta_2-\beta_1-\frac{c\Delta t}{r_0} \right)> 0.
\label{eq:calR}
\eeqa
If the interval $\Delta t$ exceeds a critical time,
$c\Delta t > (\beta_2-\beta_1)r_0$, i.e., ${\cal R}<0$,
even a photon cannot catch up with the accelerating fore shell
(see Fig.~\ref{fig:spacetime}).

The catch-up condition (\ref{eq:calR}) is not usually fulfilled
by the physical processes
because the radial velocity dispersion 
spreads the shell width
before the start $r<r_0$
and the next shell is delayed by the causal time,
$\Delta t \simg r_0/c \beta_1$ for the non-relativistic case
and $\Delta t \simg r_0/2 c \Gamma_1^2$ 
[and, hence, ${\cal R} \siml -\Gamma_1^2/(\Gamma_2^2-\Gamma_1^2)$] 
for the relativistic case,
leading to no collision.
This is the essential reason 
why we have not considered the internal shock 
in the radiation-dominated jet,
although a short timescale may be produced
by a small-scale magnetic reconnection
or a jet opening angle of less than $\sim 1/\Gamma^2$.

However, the Lorentz factor evolves differently
from the conical one $\Gamma \propto r$ to
\beqa
\Gamma \propto \left(\frac{r}{r_0}\right)^{\lambda},
\label{eq:Gamma}
\eeqa
if the jet is confined to a converging opening angle,
\beqa
\theta \propto r^{\lambda-1},
\quad (\lambda<1)
\label{eq:theta}
\eeqa
by some external pressure or internal magnetic field.
This is because the comoving volume evolves 
differently from the conical one \cite{Meszaros:2006rc} to
$V'\sim \pi \theta^2 r^2 \Gamma \Delta \propto r^{2\lambda} \Gamma$,
where the lab-frame width $\Delta$ remains constant if
it is smaller than the spreading width $\sim r/\Gamma^2$.
The adiabatic expansion of radiation 
(with an adiabatic index $\gamma_a=4/3$) decreases
the comoving temperature as 
$T' \propto V^{1-\gamma_a} \propto 
V'^{-1/3} \propto r^{-2\lambda/3} \Gamma^{-1/3}$.
A solid boundary
conserves the total energy $\Gamma T'^4 V' \propto \Gamma T' =$ const,
which yields Eq.~(\ref{eq:Gamma}).

For the confined jet $\lambda<1$, 
the catch-up condition differs from Eq.~(\ref{eq:calR}).
We can analytically integrate the trajectory in Eq.~(\ref{eq:drdt})
considering (hereafter)
the relativistic case $\Gamma_2>\Gamma_1 \gg 1$
with an expansion $\beta \simeq 1-\Gamma^{-2}/2$.
Then, the shells collide at
\beqa
\frac{r}{r_0} \simeq
\left[1-\frac{c\Delta t}{\frac{1}{2\lambda-1}\left(1-\frac{\Gamma_1^2}{\Gamma_2^2}\right)\frac{r_0}{2\Gamma_1^2}}\right]^{\frac{1}{1-2\lambda}}
\equiv {\cal R}_{\lambda}^{\frac{1}{1-2\lambda}},
\quad {\rm if}\ {\cal R}_{\lambda}>0.
\label{eq:calRl}
\eeqa
Note that ${\cal R}_{\lambda}={\cal R}$ 
for $\lambda=1$ and $\Gamma_2>\Gamma_1 \gg 1$
in Eqs.~(\ref{eq:calR}) and (\ref{eq:calRl}).

The qualitative properties change at $\lambda=1/2$.

(i) {\it Strong confinement case}, $\lambda \le 1/2$:
The collision always occurs (${\cal R}_{\lambda}>0$)
for any interval $\Delta t$, simply because
the causal region $\sim r/2 \Gamma^2 \propto r^{1-2\lambda}$ 
monotonically increases in the lab frame.
This is almost the same as the usual internal shock 
after the acceleration ceases \cite{Rees:1994nw,Kobayashi:1997jk}.
Meanwhile, since $\theta < \Gamma^{-1}$, the jet boundary is also 
causally connected with the jet axis.
This enables the recollimation or oblique shocks across the jet.
Therefore, the jet can be dissipative during the expansion
with shocks in the causal region.
Note that the strong confinement ($\theta \Gamma<1$)
is employed to accelerate
the Poynting-dominated jet to
the matter-dominated jet. \cite{Komissarov:2008ic,Tchekhovskoy:2009mf}

(ii) {\it Weak confinement case}, $1/2 < \lambda < 1$: 
The catch-up condition ${\cal R}_\lambda > 0$ in Eq.~(\ref{eq:calRl}) 
requires that the interval $\Delta t$ is less than
a critical time $\Delta t_{\rm crit}$,
\beqa
c \Delta t < \frac{1}{2\lambda-1}\left(1-\frac{\Gamma_1^2}{\Gamma_2^2}\right)\frac{r_0}{2\Gamma_1^2} \equiv c \Delta t_{\rm crit}.
\label{eq:tcrit}
\eeqa 
This can be satisfied by the causal processes (i.e., with the
next shell ejected after the causal time $\Delta t \simg
r_0/2c\Gamma_1^2$), in contrast to the conical jet ($\lambda=1$),
because the prefactor
$(2\lambda-1)^{-1}(1-\Gamma_1^2/\Gamma_2^2)$ can be larger than unity
due to the slow acceleration $\Gamma \propto r^{\lambda}$.
The space-time diagram in Fig.~\ref{fig:spacetime} is useful to
understand the following.
\begin{itemize}
\item[(A)] The radiation-dominated jet can cause internal shocks
even during the acceleration
if the jet is more collimated than the conical shape,
$\lambda<1$ in Eq.~(\ref{eq:Gamma}).
\item[(B)] The shock can occur at a large radius
since an accelerating shell asymptotically approaches a light path
(see dashed lines in Fig.~\ref{fig:spacetime}).
The shock radius is large when the interval $\Delta t$ 
gets close to the critical time $\Delta t_{\rm crit}$
where the time-fraction vanishes 
${\cal R}_\lambda=1-\Delta t/\Delta t_{\rm crit}=0$ in Eqs.~(\ref{eq:calRl}) and (\ref{eq:tcrit}).
\item[(C)] The slow shell can be repeatedly shocked 
by the successive rapid shells,
which can be ejected with shorter intervals
$\sim r_0/2c\Gamma_2^2$
than the slow shells $\sim r_0/2c\Gamma_1^2$.
\end{itemize}
In the weak confinement,
the causality is lost across the jet ($\theta \Gamma > 1$).
Although the angular structure of $\Gamma$ would arise
depending on the initial condition,
the acceleration is likely reduced
because of denser streamlines (smaller $V'$) than in the conical case.
If the confinement is strong initially (inside the star) and weak later 
(outside the star), the oblique shock does not necessarily 
accompany the weak confinement phase.

~

\noindent {\it \S Jet confinement by cocoon} --
The jet would be confined by the progenitor star (long GRB)
or the surrounding dense matter or wind \cite{Shibata:2011fj}
(short GRB)
\cite{Begelman:1989,Matzner:2002ti,Bromberg:2007kw,RamirezRuiz:2002jq}.
Note that the merger of two neutron stars ejects 
baryons before the collapse 
to a black hole even with an ordinary magnetic field 
strength.\cite{Shibata:2011fj}
As the jet runs into the stellar envelope,
the shocked jet and envelope flow sideways (if $\theta<\Gamma_h^{-1}$)
and inflate a cocoon, whose pressure confines the jet.
We can roughly estimate the index $\lambda$ 
in Eqs.~(\ref{eq:Gamma}) and (\ref{eq:theta})
by considering the pressure balance at 
\begin{itemize}
\item[(I)] the jet head: The longitudinal balance of the ram pressure between
the jet $\sim L_j/\pi \theta^2 r^2 c$
and the stellar matter $\sim \rho_* c^2 \beta_h^2$
determines the jet head velocity
$c \beta_h \sim c (L_j/\pi c^3 \rho_* r^2)^{1/2} \theta^{-1}$.
\item[(II)] the cocoon-envelope boundary: 
The cocoon expands with a speed $c\beta_c$
by balancing the ram pressure of the ambient matter 
$\sim \rho_* c^2 \beta_c^2$
and the cocoon pressure $\sim L_j t/(c\beta_h t)(c\beta_c t)^2$
which is the total energy deposited by the jet
divided by the volume of the cocoon,
so that $\beta_c \sim (L_j/c^3 \rho_* r^2)^{1/4} \beta_h^{1/4} \sim (L_j/c^3 \rho_* r^2)^{3/8} \theta^{-1/4}$
with $r \sim c \beta_h t$.
\item[(III)] the jet-cocoon boundary: 
The cocoon pressure $\sim \rho_* c^2 \beta_c^2$ confines the 
radiation-dominated (hot) jet 
with the transverse pressure $p_j \sim L_j/\pi \theta^2 r^2 c \Gamma^2$,
leading to $\theta \sim (L_j/c^3 \rho_* r^2)^{1/6} \Gamma^{-4/3}$.
\end{itemize}
Substituting the envelope profile $\rho_* \sim r^{-n}$,
$\Gamma \propto r^{\lambda}$ in Eq.~(\ref{eq:Gamma}),
$\theta \propto r^{\lambda-1}$ in Eq.~(\ref{eq:theta}), and 
a constant luminosity $L_j \sim$ const,
we can equate the radial dependence to find
\beqa
\lambda \sim \frac{4+n}{14}. \quad 
\left(\theta \propto r^{(n-10)/14}\ {\rm inside\ the\ star}\ 
\rho_* \sim r^{-n}\right)
\label{eq:lambda1}
\eeqa
The ideal radiative envelope has an index $n=3$
(i.e., $\lambda \sim 1/2$),
whereas more realistic presupernova models such as 16TI \cite{Woosley:2005gy}
have shallower slopes $n \sim 2$
(i.e., more confinement $\lambda \sim 3/7$).
Because of the strong confinement $\lambda \le 1/2$, 
the jet is basically dissipative before the jet breakout.
Similar dissipation is observed in the simulations.
\cite{Aloy:1999ai,Mizuta:2004gu,Lazzati:2009xx}
The Lorentz factor of the jet at the breakout is around 
\beqa
\Gamma_b \sim \left(\frac{r_*}{r_{\rm ISCO}}\right)^{\lambda}
\sim 10{\rm -}30,
\label{eq:Gammab}
\eeqa
where $r_* \sim 10^{10}$ cm is the stellar radius (breakout radius)
and $r_{\rm ISCO} \sim 10^7$ cm is the central engine size.
The variability timescale is about 
$\sim r_*/c \Gamma_b^2$, which is $\sim r_{\rm ISCO}/c$ for $\lambda=1/2$,
so that the central engine timescale may be marginally preserved.

Even after the jet breakout,
the jet would be confined to some extent.
A plausible source of pressure is the fast cocoon 
that just escapes from the jet head sideways
at the breakout.
The fast cocoon is thereby fast (ranging from $\Gamma \sim 1$) 
up to the jet Lorentz factor
at the breakout $\sim \Gamma_b$ in Eq.~(\ref{eq:Gammab}),
and confines the following jet with a lag time $\delta t$
over a long distance 
$\sim c \delta t \Gamma_b^2 \sim 10^{13} 
(\delta t/1\ {\rm s}) (\Gamma_b/30)^2$ cm.
Also, the transverse pressure of the fast cocoon is initially comparable to
that of the jet $p_j$,
and then declines slower 
$p_c \propto T'^4 \propto V'^{-4/3} \propto r^{-8/3}$
than 
$p_j \sim L_j/\pi \theta^2 r^2 c \Gamma^2 \propto r^{-4\lambda}$
if the jet were conical ($\lambda=1$),
because the fast cocoon becomes matter-dominated\footnote{
The fast cocoon may be initially radiation-dominated
and expand to $\Gamma \sim \gamma_p' \Gamma_b \sim \Gamma_b^2$ after
converting the radiation energy into the bulk motion,
where $\gamma_p' \sim \Gamma_b$ is 
the random Lorentz factor per proton at the shock,
as in the collisionless bulk acceleration. \cite{Ioka:2010xc}
Then, the Lorentz factor of the fast cocoon may range 
from $\Gamma \sim 1$ to $\sim \Gamma_b^2$.
This effect could alter $\lambda$ in Eq.~(\ref{eq:lambda2}).
}
with the coasting expansion $V' \sim 4\pi r^2 \Gamma \Delta \propto r^2$.
For $p_c \sim p_j$, the jet is confined,
\beqa
\lambda \sim \frac{2}{3} . \quad (\theta \propto r^{-1/3}\ {\rm outside\ the\ star})
\label{eq:lambda2}
\eeqa
The confinement gets weaker than inside the star
(with index larger than the boundary $\lambda > 1/2$)
but still continues (in the weak regime $1/2< \lambda < 1$).
Thus, the dissipation proceeds via the internal shocks
caused by the variability due to the central engine or 
the jet-envelope interaction.
The initial opening angle would be 
the inverse of the Lorentz factor 
$\theta_0 \sim \Gamma_b^{-1} \sim 0.03$--$0.1$
in Eq.~(\ref{eq:Gammab}) 
at the stellar surface $r_0 \sim r_* \sim 10^{10}$ cm,
and the product $\Gamma \theta \propto r^{2\lambda-1}$ grows afterward,
as required for the afterglow observations of jet breaks.

\begin{figure}[t]
%\centerline{\includegraphics[scale=.32]{f1.eps}}
\centerline{\includegraphics[scale=.3]{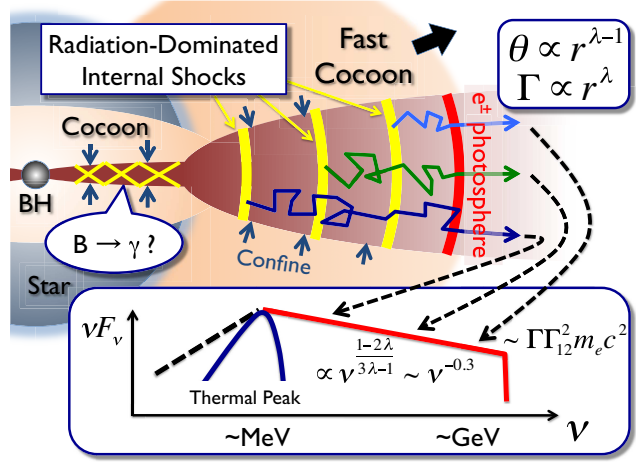}}
\caption{
Radiation-dominated internal shocks (RDISs) can occur 
if the jet is confined in $\theta \propto r^{\lambda-1}$ with $\lambda<1$,
for example, by the fast cocoon pressure [Eq.~(\ref{eq:lambda2})],
because the jet acceleration slows down $\Gamma \propto r^{\lambda}$
[Eq.~(\ref{eq:Gamma})].
The RDISs accelerate photons and create $e^{\pm}$.
The nonthermal photons are generated from extended radii,
where lower energy photons originate deeper with more scatterings
[Eqs.~(\ref{eq:tau}) and (\ref{eq:nuth})]
before emerging from the $e^{\pm}$ photosphere
[Eq.~(\ref{eq:rpm})].
In this paper, we do not account for the low-energy spectral slope.
}
\label{fig:spec}
\end{figure}

~

\noindent {\it \S $e^{\pm}$ photosphere and high $\Gamma$ with internal shocks} --
When the rapid jet catches up with the slow one,
an internal shock forms, which dissipates 
the relative kinetic energy.
The relative Lorentz factor at the collision 
is the same as the initial value,
\beqa
\Gamma_{12} \simeq \frac{1}{2} 
\left(\frac{\Gamma_2}{\Gamma_1}+\frac{\Gamma_1}{\Gamma_2}\right),
\eeqa
since $\Gamma \propto r^{\lambda}$ for both jets.
For the radiation-dominated jet,
the kinetic energy is almost carried by photons,
which is transferred
by Compton scatterings of $e^{\pm}$,
forming a radiation-mediated shock \cite{Budnik:2010ru}.
A small fraction of photons are repeatedly upscattered 
across the shock,
extending a power-law spectrum to high energies
like the Fermi acceleration 
(Blandford-Payne mechanism)
\cite{BP:1981,Wang:2006jc}.
Even for a moderate optical depth, $\tau_T \sim$ a few,
the Compton $y$-parameter $y \sim \tau_T \beta_{12}^2 \Gamma_{12}^2$
can exceed unity for the relativistic shocks.
After the shock passage,
the turbulent motion 
induced by, for example, Richtmyer-Meshkov instability,
also provides the scattering centers, continuing 
the photon acceleration. \cite{Inoue:2010eu,Thompson:1994}
Then, an appreciable fraction of the relative kinetic energy
can go to high-energy photons and create $e^{\pm}$ pairs.

The $e^{\pm}$ pair creation can continue to a large radius
since the collision radius rapidly grows as 
the jets expand owing to radiation pressure
[see Eqs.~(\ref{eq:calRl}) and (\ref{eq:tcrit}) and Fig.~\ref{fig:spacetime}].
The jet acceleration holds
as long as the $e^{\pm}$ pairs trap the radiation.
The maximum Lorentz factor is limited by the condition 
that the system becomes transparent
$\tau_T \sim n'_{\pm} \sigma_T c t'_{\rm dyn} \sim 1$,
where $n'_{\pm}=f_{\pm} L/4\pi r^2 m_e c^3 \Gamma^2$ is
the comoving number density of $e^{\pm}$ pairs,
$t'_{\rm dyn} \sim r/c \Gamma$ is the comoving dynamical time,
$L$ is the isotropic total luminosity and
$f_{\pm}$ is the energy fraction of $e^{\pm}$ pairs.
Then, the coasting Lorentz factor is
\beqa
\Gamma_{\pm}=\left(\frac{\Gamma_0^{1/\lambda} f_{\pm} L\sigma_T}{4\pi m_e c^3 r_0}\right)^{\frac{\lambda}{3\lambda+1}}
\simeq 2900 \left(\frac{\Gamma_0}{30}\right)^{1/3}
\left(\frac{r_0}{10^{10}\ {\rm cm}}\right)^{-2/9}
\left(\frac{f_{\pm} L}{10^{53}\ {\rm erg}\ {\rm s}^{-1}}\right)^{2/9},
\eeqa
at the $e^{\pm}$ photosphere of radius 
\beqa
r_{\pm} \sim r_0 \left(\frac{\Gamma_{\pm}}{\Gamma_0}\right)^{1/\lambda} 
\sim 1\times 10^{13}\ {\rm cm}\ 
\left(\frac{\Gamma_0}{30}\right)^{-1}
\left(\frac{r_0}{10^{10}\ {\rm cm}}\right)^{2/3}
\left(\frac{f_{\pm} L}{10^{53}\ {\rm erg}\ {\rm s}^{-1}}\right)^{1/3},
\label{eq:rpm}
\eeqa
where (and hereafter) we suppose
a jet breaking out of the stellar surface,
$r_0 \sim r_* \sim 10^{10}$ cm
and $\Gamma_0 \sim \Gamma_b \sim 10$--$30$
in Eq.~(\ref{eq:Gammab}),
with $\lambda=2/3$ in Eq.~(\ref{eq:lambda2})
in the last equalities.
The pair energy fraction $f_{\pm}$
is about
the energy fraction
emitted above the pair creation threshold $\sim \Gamma_{\pm} m_e c^2 \sim$ GeV,
and $f_{\pm}\sim 1$ is implied in Fermi bursts.
The high Lorentz factor $\Gamma_{\pm} > 10^3$
is also consistent with the Fermi observations.

~

\noindent {\it \S Nonthermal photospheric spectrum} --
The radiation-dominated jet emits almost all the energy
from the $e^{\pm}$ photosphere,
which is nonthermalized by internal shocks.
The spectrum would have a thermal peak with a high-energy power-law tail.
We illustrate the spectral shaping
using a constant-luminosity model:
equal-energy rapid jets are ejected after a slow jet
at intervals of the causal time
$\Delta t_{\rm sh} \sim r_0/2c \Gamma_2^2$ $(\ll r_0/2c \Gamma_1^2)$ 
(see Fig.~\ref{fig:spacetime}).
The internal shocks can repeat 
in the dynamical time of the slow jet
and create sufficient $e^{\pm}$ pairs to sustain the opacity.
Thus, the $e^{\pm}$ pair density 
is mainly determined by the $e^{\pm}$ annihilation
as $n'_{\pm} \sim 1/\sigma_T c \Delta t'_{\rm sh}$
after a time $\Delta t'_{\rm sh}$ 
following a shock.
The interval of shocks is redshifted in
the comoving frame as
$\Delta t'_{\rm sh} \sim \Gamma \Delta t_{\rm sh}
\sim \Gamma r_0/2c \Gamma_2^2$.
Then, the typical optical depth of the jet evolves as
\beqa
\tau_T \sim n'_{\pm} \sigma_T c t'_{\rm dyn}
\sim \frac{t'_{\rm dyn}}{\Delta t'_{\rm sh}}
\sim \frac{r/c\Gamma}{\Gamma r_0/2c\Gamma_2^2}
\propto \frac{r}{\Gamma^2} \propto r^{1-2\lambda} \sim r^{-1/3}.
\label{eq:tau}
\eeqa
Note that $\tilde \tau_T \propto \tilde n'_{\pm} t'_{\rm dyn}
\propto V'^{-1} r/\Gamma
\propto r^{-4\lambda+1} \propto r^{-5/3}$
(steeper)
if $e^{\pm}$ were conserved.

Under the photosphere $\tau_T>1$,
thermal photons carry the dominant energy
at a (nonrelativistic) comoving temperature $h \nu'_{\rm peak} < m_e c^2$
with a number density
$n'_{\rm peak} \sim f_{\pm}^{-1} n'_{\pm} m_e c^2/h \nu'_{\rm peak}$.
In this cool bath, $e^{\pm}$ cools by
$\Delta T'_{e}/T'_{e} \sim -h \nu'_{\rm peak}/m_e c^2$
in a single Compton scattering,
and becomes thermal 
in a time less than the dynamical time,
$t'_{{\rm cool},e}
\sim (n'_{\rm peak} \sigma_T c)^{-1} |T'_{e}/\Delta T'_{e}|
\sim t'_{\rm dyn} f_{\pm}/\tau_T < t'_{\rm dyn}$.
The nonthermal photons with $\nu'>\nu'_{\rm peak}$
also lose energy
$\Delta \nu'/\nu' \sim (4kT'_e-h \nu')/m_ec^2 \sim -h \nu'/m_e c^2$
in a single scattering by the cooled $e^{\pm}$.
However, the cooling time
exceeds the dynamical time,
$t'_{{\rm cool},\gamma} \sim 
(n'_{\pm}\sigma_T c)^{-1}|\nu'/\Delta \nu'| > t'_{\rm dyn}$,
at $h \nu'< m_e c^2/\tau_T$.
Therefore,
the spectrum remains nonthermal below the observed frequency
$\sim \Gamma \nu'$ ($\sim$ const),
\beqa
\nu \sim \frac{\Gamma m_e c^2}{h \tau_T} \propto r^{3\lambda-1} 
\sim r^{1},
\label{eq:nuth}
\eeqa
even if the photons are generated below the photosphere $r<r_{\pm}$
with $\tau_T>1$.
At $\tau_T>1$, the photons are trapped
since the diffusion time is 
$t_{\rm diff} \sim \tau_T^2/n'_{\pm}\sigma_T c 
\sim \tau_T t_{\rm dyn}$.

The internal shock extends the spectrum to a flat power law,
injecting a fair fraction of the shock energy 
at the frequency in Eq.~(\ref{eq:nuth}).
In the constant-luminosity model, 
the injected shock energy is proportional to the number of shocks
$\sim t'_{\rm dyn}/\Delta t'_{\rm sh}$,
so that the emergent spectrum at the photosphere is
\beqa
\nu F_{\nu} \propto \frac{t'_{\rm dyn}}{\Delta t'_{\rm sh}} 
\propto r^{1-2\lambda}
\propto \nu^{(1-2\lambda)/(3\lambda-1)} \sim \nu^{-0.3},
\label{eq:spec}
\eeqa
with Eqs.~(\ref{eq:tau}) and (\ref{eq:nuth}).
Interestingly, this is consistent with
the high-energy Band spectrum of GRBs 
$\nu F_{\nu} \propto \nu^{-0.3 \pm 0.3}$. \cite{Meszaros:2006rc,Zhang:2007nka}
The nonthermal photons are generated from radii extending
over several orders,
where the lower energy photons originate deeper below the photosphere
with more scatterings
in Eqs.~(\ref{eq:tau}) and (\ref{eq:nuth}).

Our model naturally explains the observations that
the nonthermal energy above $\nu_{\rm peak}$
is comparable to the thermal energy
because the radiation itself nonthermalizes the radiation.
In the constant-luminosity model,
the slow jets (mainly for thermal energy) and 
rapid jets (mainly for nonthermal energy)
have comparable energies
if they are ejected
for a similar duration
of the causal time of the slow jets
$\sim r_0/2c\Gamma_1^2$
(see Fig.~\ref{fig:spacetime}).

The rapid jets could become transparent via $e^{\pm}$ annihilation
before colliding with the slow jet.
The released photons expand conically and
only a part of them $\sim (\theta/\theta_0)^2 \propto r^{2\lambda-2}$
could shock the slow jet.
The loss of shocked energy softens the spectrum as
$\nu F_{\nu} \propto r^{1-2\lambda} \times r^{2\lambda-2} 
\propto \nu^{1/(1-3\lambda)} \sim \nu^{-1}$
with Eqs.~(\ref{eq:nuth}) and (\ref{eq:spec}),
which is still consistent with the observations.
The angular structure of $\lambda$ [below Eq.~(\ref{eq:tcrit})]
could also cause diversity to the high-energy index.
The released photons could be absorbed by
the rear jets with different streamlines,
or by the surrounding (optically thick) fast cocoon,
possibly developing the hollow cone-structured jet 
with $\Gamma \sim \Gamma_b \sim 10$--$30$
for the shallow X-ray afterglow
(see below).
The decoupled $e^{\pm}$ from the rapid jet could be caught up with
by other rear jets and create further $e^{\pm}$.

~

\noindent {\it \S Implications} --

\noindent $\bullet$ {\it GeV-TeV spectrum and delay} --
The photospheric spectrum may extend to
$\nu_{\max} \sim \Gamma_{\pm} \Gamma_{12}^2 m_e c^2 
\sim 10$ GeV$(\Gamma_{\pm}/3000) (\Gamma_{12}/3)^2$,
\cite{Budnik:2010ru}
where we expect the $e^{\pm}$ creation cutoff and 
possibly the (blue-shifted and broadened) annihilation line,
\cite{Pe'er:2003ft,Ioka:2007qk}
providing closure relations \cite{Murase:2007ya}
for Fermi and CTA
to verify the $e^{\pm}$ photosphere.
We may also expect a spectral break at 
$\nu \sim \Gamma_{\pm} \Gamma_{12} m_e c^2$
due to the Klein-Nishina effect.

The GeV onset delay observed in Fermi bursts
\cite{Abdo:2009a,Ackermann:2010,Zhang:2010ey}
may be explained by
the leading jet that is not confined by the fast cocoon
and, hence, cannot keep internal shocks and $e^{\pm}$ 
to a high Lorentz factor.
In this picture, the delay time is about
$\sim r_{\pm}/c \Gamma_b^2 \sim 1$ s 
in Eqs.~(\ref{eq:Gammab}) and (\ref{eq:rpm}), as observed.

The (collisionless) internal shocks
continue beyond the photosphere 
within $e^{\pm}$ outflows
after the radiation decoupling,
and could produce the observed extra GeV component,
possibly up to TeV, by
synchrotron or inverse Compton emission
\cite{Ioka:2010xc,Toma:2010xw}.
The luminosity can be appreciable up to $\sim f_{\pm} L$.

\noindent $\bullet$ {\it Early afterglow} --
The prompt emission is radiatively efficient
$1-f_{\pm} \simg 50\%$,
and the remaining kinetic energy of $e^{\pm}$ powers the early afterglow.
The $e^{\pm}$ energy fraction $f_{\pm}$ 
can be read from the prompt spectrum
above the pair creation threshold
$\sim \Gamma_{\pm} m_e c^2 \sim$ GeV.
Thus, we predict
a correlation between the early afterglow and the GeV-TeV prompt emission:
a steep decay of X-ray afterglow \cite{Zhang:2005fa,Ioka:2005zj} 
accompanies a low GeV-TeV energy fraction.
Further Fermi-Swift co-observations would be helpful.
The shallow X-ray phase \cite{Zhang:2005fa,Ioka:2005zj}
could be produced by
the (matter-dominated) fast cocoon with $\Gamma \sim 10$--$30$
that receives energy via photons from $e^{\pm}$-annihilated jets
[below Eq.~(\ref{eq:spec})].
The reverse shock emission would be soft because of
the increased number of emitting leptons.

\noindent $\bullet$ {\it Cosmic Rays} --
The RDISs would produce
few ultra high-energy cosmic rays
and neutrinos, \cite{Waxman:1997ti}
consistent with the IceCube upper limit. \cite{Abbasi:2011qc}
The 
fast cocoon energized by photons from $e^{\pm}$-annihilated jets
might contribute to
these emissions.

\noindent $\bullet$ {\it Central engine} --
The RDIS model
requires less fine tuning of baryon load than previously thought,
expanding the parameter space even to 
the original pure-leptonic model. \cite{p86,g86}
The baryon load could be prevented by the dipole field
near a black hole, which may be rather necessary for
the Blandford-Znajek effect to launch a relativistic jet
despite nonaxisymmetric turbulence. \cite{McKinney:2008ev}
The ejected Poynting-dominated jet is strongly confined
in the star or surrounding matter,
possibly dissipating into the radiation-dominated jet
via turbulence.

\noindent $\bullet$ {\it Unresolved problems} --
We leave 
the low-energy spectral index \cite{Meszaros:2006rc,Zhang:2007nka}
$\nu F_{\nu} \propto \nu$
and the Amati and Yonetoku spectral relations 
\cite{Amati:2002ny,Yonetoku:2003gi}
for future discussion.
Since $\Gamma T'=$ const, the peak energy reflects the initial fireball
temperature.
%We are also calculating the transverse structure of the jet
%by numerical simulations.

\section*{Acknowledgements}
We thank a referee,
K.~Asano, A.~Beloborodov, J.~C.~McKinney, T.~Nakamura and T.~Piran 
for useful comments.
%discussions.
%a careful reading of the manuscript.
%We also acknowledge helpful comments and suggestions from
%an anonymous referee.
This work is supported
%in part
by KAKENHI,
%the Grant-in-Aid from the 
%Ministry of Education, Culture, Sports, Science and Technology
%(MEXT) of Japan, 
%Nos. 
19047004, 22244019, 22244030 (KI),
21684014 (KI, YO),
22740131 (NK) and 20105005 (AM).

%\appendix
%\section{First Appendix} %Empty argument \section{} yields `Appendix'. 
%
%\section{Second Appendix}

\end{document}